\documentclass[12pt]{article}
\usepackage{amsmath}
\usepackage{amssymb}

\input{tcilatex}

\begin{document}

\title{\textbf{General Relativity and Weyl Geometry }}
\author{C. ROMERO,\thanks{%
cromero@fisica.ufpb.br}{\ } J. B. FONSECA-NETO and M. L. PUCHEU}
\maketitle

\begin{abstract}
We show that the general theory of relativity may be formulated in the
language of Weyl geometry. We develop the concept of Weyl frames and point
out that the new mathematical formalism may lead to different pictures of
the same gravitational phenomena. We show that in an arbitrary Weyl frame
general relativity, which takes the form of a scalar-tensor gravitational
theory, is invariant with respect to Weyl tranformations. A kew point in the
development of the formalism is to build an action that is manifestly
invariant with respect to Weyl transformations. When this action is
expressed in terms of Riemannian geometry we find that the theory has some
similarities with Brans-Dicke gravitational theory. In this scenario, the
gravitational field is not described by the metric tensor only, but by a
combination of both the metric and a geometrical scalar field. We illustrate
this point by, firstly, discussing the Newtonian limit in an arbitrary
frame, and, secondly, by examining how distinct geometrical and physical
pictures of the same phenomena may arise in different frames. To give an
example, we discuss the gravitational spectral shift as viewed in a general
Weyl frame. We further explore the analogy of general relativity with
scalar-tensor theories and show how a known Brans-Dicke vacuum solution may
appear as a solution of general relativity theory when reinterpreted in a
particular Weyl frame. Finally, we show that the so-called WIST gravity
theories are mathematically equivalent to Brans-Dicke theory when viewed in
a particular frame.

\end{abstract}

keywords: {Weyl frames; conformal transformations; general relativity.}


address: {Departamento de F\'{\i}sica, Universidade Federal da Para\'{\i}ba,
Jo\~{a}o Pessoa, PB 58059-970, Brazil}

\section{Introduction}

We would like to start by raising two questions of a very general character.
The first questions is: What kind of invariance should the basic laws of
physics possess? It is perhaps pertinent here to quote the following words
by Dirac: \textquotedblleft It appears as one of the fundamental principles
of nature that the equations expressing the basic laws of physics should be
invariant under the widest possible group of
transformations\textquotedblright\ \cite{Dirac}. The second question, which
seems to be of a rather epistemological character, is: To what extent is
Riemannian geometry the only possible geometrical setting for general
relativity? The purpose of the present work is to address, at least
partially, these two questions.

It is a very\ well known fact that the principle of general covariance has
played a major role in leading Einstein to the formulation of the theory of
general relativity \cite{Norton}. The idea underlying this principle is that
coordinate systems are merely mathematical constructions to\ conveniently
describe physical phenomena, and hence should not be an essential part of
the fundamental laws of physics. In a more precise mathematical language,
what is being required is that the equations of physics be expressed in
terms of intrinsic geometrical objects, such as scalars, tensors or spinors,
defined in the space-time manifold. This mathematical requirement is
sufficient to garantee the invariance of the form of the physical laws (or
covariance of the equations) under arbitrary coordinate transformations. In
field theories, one way of constructing covariant equations is to start with
an action in which the Lagrangian density is a scalar function of the
fields. In the case of general relativity, as we know, the covariance of the
Einstein equations is a direct consequence of the invariance of the
Einstein-Hilbert action with respect to space-time diffeomorphisms.

A rather different kind of invariance that has been considered in some
branches of physics is invariance under conformal transformations. These
represent changes in the units of length and time that differ from point to
point in the space-time manifold. Conformal transformations were first
introduced in physics by H. Weyl in his attempt to formulate a unified
theory of gravitation and electromagnetism \cite{Weyl}. However, in order to
introduce new degrees of freedom to account for the electromagnetic field
Weyl had to assume that the space-time manifold is not Riemannian. This
extension consists of introducing an extra geometrical entity in the
space-time manifold, a 1-form field $\sigma $, in terms of which the
Riemannian compatibility condition between the metric $g$ and the connection 
$\nabla $ is redefined. Then, a group of transformations, which involves
both $g$ and $\sigma $,\ is defined by requiring that under these
transformations the new compatibility condition remain invariant. In a
certain sense, this new invariance group, which we shall call the group of
Weyl transformations, includes the conformal transformations as subgroup.

It turns out that Einstein's theory of gravity in its original formulation
is not invariant neither under conformal transformations nor under Weyl
transformations. One reason for this is that the geometrical language of
Einstein's theory is completely based on Riemannian geometry. Indeed, for a
long time general relativity has been inextricably associated with the
geometry of Riemann. Further developments, however, have led to the
discovery of different geometrical structures, which we might \ generically
call \textquotedblleft non-Riemannian" geometries, Weyl geometry being one
of the first examples. Many of these developments were closely related to
attempts at unifying gravity and electromagnetism \cite{Goenner}. While the
newborn non-Riemannian geometries were invariably associated with new
gravity theories, one question that naturally arises is to what extent is
Riemannian geometry the only possible geometrical setting for the
formulation of general relativity. One of our aims in this paper is to show
that, surprisingly enough, one can formulate general relativity using the
language of a non-Riemannian geometry, namely, the one known as Weyl
integrable geometry. In this formulation, general relativity appears as a
theory in which the gravitational field is described simultaneously by two
geometrical fields: the metric tensor and the Weyl scalar field, the latter
being an essential part of the geometry, manifesting its presence in almost
all geometrical phenomena, such as curvature, geodesic motion, and so on. As
we shall see, in this new geometrical setting general relativity exhibits a
new kind of invariance, namely, the invariance under Weyl transformations.

The outline of this paper is as follows. We begin by presenting the basic
mathematical facts of Weyl geometry and the concept of Weyl frames. In
section~3, we show how to recast general relativity in the language of Weyl
integrable geometry. In this formulation, we shall see that the theory is
manifestly invariant under the group of Weyl transformations. We proceed, in
section~4, to obtain the field equations and interpret the new form of the
theory as a kind of scalar-tensor theory of gravity. In sections~5 and 6, \
we explore the similarities of the formalism with Brans-Dicke theory of
gravity. We devote section 7 to examine the Newtonian limit to get some
insight into the meaning of the scalar field in the Weyl representation of
general relativity. Then, in section 8, we briefly illustrate how different
pictures of the same phenomena may arise in distinct frames. In section 9,
we show that the so-called WIST gravity theories are mathematically
equivalent to Brans-Dicke theory when viewed in a particular frame, the
Riemann frame. We end up with some remarks in section 10.

\section{Weyl Geometry}

The geometry conceived by Weyl is a simple generalization of Riemannian
geometry. Instead of postulating that the covariant derivative of the metric
tensor $g$ is zero, we assume the more general condition \cite{Weyl}

\begin{equation}
\nabla _{\alpha }g_{\mu \nu }=\sigma _{\alpha }g_{\mu \nu },\text{ }
\label{compatibility}
\end{equation}%
where $\sigma _{\alpha }$ \ denotes the components with respect to a local
coordinate basis $\left\{ \frac{\partial }{\partial x^{\alpha }}\right\} $
of a one-form field $\sigma \ $defined on the manifold $M$. This represents
a generalization of the Riemannian condition of compatibility between the
connection $\nabla $ and $g,$ namely, the requirement that the length of a
vector remain unaltered by parallel transport \cite{Pauli}. If $\sigma $
vanishes, then (\ref{compatibility}) reduces to the familiar Riemannian
metricity condition. It is interesting to note that the Weyl condition (\ref%
{compatibility}) remains unchanged\ when we perform the following
simultaneous transformations in $g$ and $\phi $:%
\begin{equation}
\overline{g}=e^{f}g,  \label{conformal}
\end{equation}%
\begin{equation}
\overline{\sigma }=\sigma +df,  \label{gauge1}
\end{equation}%
where $f$ is a scalar function defined on $M$. If $\sigma $ $=d\phi ,$ where 
$\phi $ is a scalar field, then we have what is called\ a\textit{\ Weyl
integrable manifold}. The set $(M,g,\phi )$ consisting of a differentiable
manifold $M$ endowed with a metric $g$ and a Weyl scalar field $\phi $ $\ $%
will be referred to as a \textit{Weyl frame}. In the particular case of a
Weyl integrable manifold (\ref{gauge1})\ becomes 
\begin{equation}
\overline{\phi }=\phi +f.  \label{gauge}
\end{equation}

It turns out that if the Weyl connection $\nabla $ is assumed to be
torsionless, then by virtue of condition (\ref{compatibility}) it gets
completely determined by $g$ and $\sigma $. Indeed, a straightforward
calculation shows that the components of the affine connection with respect
to an arbitrary vector basis completely are given by%
\begin{equation}
\Gamma _{\mu \nu }^{\alpha }=\{_{\mu \nu }^{\alpha }\}-\frac{1}{2}g^{\alpha
\beta }[g_{\beta \mu }\sigma _{\nu }+g_{\beta \nu }\sigma _{\mu }-g_{\mu \nu
}\sigma _{\beta }],  \label{Weylconnection}
\end{equation}%
where $\{_{\mu \nu }^{\alpha }\}=$ $\frac{1}{2}g^{\alpha \beta }[g_{\beta
\mu ,\nu }+g_{\beta \nu ,\mu }-g_{\mu \nu ,\beta }]$ represents the
Christoffel symbols, i.e., the components of the Levi-Civita connection 
\footnote{%
Throughout this paper our convention is that Greek indices take values from $%
0$ to $n-1$, where $n$ is the dimension of $M.$}. An important fact that
deserves to be mentioned is the invariance of the affine connection
coefficients $\Gamma _{\mu \nu }^{\alpha }$ under the Weyl transformations (%
\ref{conformal}) and (\ref{gauge1}). If $\sigma $ $=d\phi $, (\ref%
{Weylconnection}) becomes 
\begin{equation}
\Gamma _{\mu \nu }^{\alpha }=\{_{\mu \nu }^{\alpha }\}-\frac{1}{2}g^{\alpha
\beta }[g_{\beta \mu }\phi ,_{\nu }+g_{\beta \nu }\phi ,_{\mu }-g_{\mu \nu
}\phi ,_{\beta }].  \label{Weyl connection1}
\end{equation}

A clear geometrical insight on the properties of Weyl parallel transport is
given by the following proposition: Let $M$ be a differentiable manifold
with an affine connection $\nabla $, a metric $g$ and a Weyl field of
one-forms $\sigma $. If $\nabla $ is compatible with $g$ in the Weyl sense,\
i.e. if (\ref{compatibility}) holds, then for any smooth curve $C=C(\lambda
) $ and any pair of two parallel vector fields $V$ and $U$ along $C,$ we
have 
\begin{equation}
\frac{d}{d\lambda }g(V,U)=\sigma (\frac{d}{d\lambda })g(V,U),
\label{covariantderivative}
\end{equation}%
where $\frac{d}{d\lambda }$ denotes the vector tangent to $C$ and $\sigma (%
\frac{d}{d\lambda })$ indicates the aplication of the 1-form $\sigma $ on $%
\frac{d}{d\lambda }$. (In a coordinate basis, putting $\frac{d}{d\lambda }=%
\frac{dx^{\alpha }}{d\lambda }\frac{\partial }{\partial x^{\alpha }},$ $%
V=V^{\beta }\frac{\partial }{\partial x^{\beta }},U=U^{\mu }\frac{\partial }{%
\partial x^{\mu }},\sigma =\sigma _{\nu }dx^{\nu },$ the above equation
reads $\frac{d}{d\lambda }(g_{\alpha \beta }V^{\alpha }U^{\beta })=\sigma
_{\nu }\frac{dx^{\nu }}{d\lambda }g_{\alpha \beta }V^{\alpha }U^{\beta }.$)

If we integrate the equation (\ref{covariantderivative})\ along the curve $C$
from a point $P_{0}=C(\lambda _{0})$ to an arbitrary point\ $P=C(\lambda ),$
then we obtain%
\begin{equation}
g(V(\lambda ),U(\lambda ))=g(V(\lambda _{0}),U(\lambda
_{0}))e^{\int_{\lambda _{0}}^{\lambda }\sigma (\frac{d}{d\rho })d\rho }.
\label{integral}
\end{equation}%
If we put $U=V$ and denote by $L(\lambda )$ the length of the vector $%
V(\lambda )$ at $P=C(\lambda )$, then it is easy to see that in a local
coordinate system $\left\{ x^{\alpha }\right\} $ the equation (\ref%
{covariantderivative}) reduces to 
\begin{equation*}
\frac{dL}{d\lambda }=\frac{\sigma _{\alpha }}{2}\frac{dx^{\alpha }}{d\lambda 
}L.
\end{equation*}

Consider the set of all closed curves $C:[a,b]\in R\rightarrow M$, i.e, with 
$C(a)=C(b).$ Then, \ we have the equation 
\begin{equation*}
g(V(b),U(b))=g(V(a),U(a))e^{\doint \sigma (\frac{d}{d\lambda })d\lambda }.
\end{equation*}%
It follows from Stokes' theorem that if $\sigma $ is an exact form, that is,
\ if there exists a scalar function $\phi $, such that $\sigma =d\phi $, then

\begin{equation*}
\oint \sigma (\frac{d}{d\lambda })d\lambda =0
\end{equation*}%
for any loop. In this case the integral $e^{\int_{\lambda _{0}}^{\lambda
}\sigma (\frac{d}{d\rho })d\rho }$ does not depend on the path and (\ref%
{integral}) may be rewritten in the form 
\begin{equation}
e^{-\phi (x(\lambda ))}g(V(\lambda ),U(\lambda ))=e^{-\phi (x(\lambda
_{0}))}g(V(\lambda _{0}),U(\lambda _{0})).  \label{isometry}
\end{equation}%
This equation means that we have an isometry between the tangent spaces of
the manifold at the points $P_{0}=C(\lambda _{0})$ and $P=C(\lambda )$ in
the "effective" metric $\widehat{g}=e^{-\phi }g$.

Let us have a closer look at the correspondence between the Riemannian and
Weyl integrable geometries suggested by Eq. (\ref{isometry}). The first
point to note is that, because $\sigma =d\phi $\ for some scalar field $\phi 
$, then if we define an "effective" metric $\widehat{g}=e^{-\phi }g$,\ the
Weyl condition of compatibility (or, as it is sometimes called, the
non-metricity condition), expressed by Eq. (\ref{compatibility}) or (\ref%
{covariantderivative}), is formally equivalent to the Riemannian condition
imposed on $\widehat{g}$, namely, 
\begin{equation*}
\nabla _{\alpha }\widehat{g}_{\mu \nu }=0.
\end{equation*}%
\ It may be easily verified that (\ref{Weyl connection1}) follows directly
from $\nabla _{\alpha }\widehat{g}_{\mu \nu }=0$. This simple fact has
interesting and useful consequences, and later will serve as a guidance in
the formulation of general relativity in terms of Weyl integrable geometry.
One consequence is that since $\widehat{g}=e^{-\phi }g$ is invariant under
the Weyl transformations (\ref{conformal}) and (\ref{gauge}) any geometrical
quantity constructed with and solely with $\widehat{g}$ \ is invariant.
Clearly, these will also be invariant under the Weyl transformations (\ref%
{conformal}) and (\ref{gauge}). Thus, in addition to the connection
coefficients $\widehat{\Gamma }_{\mu \nu }^{\alpha }=$ $\Gamma _{\mu \nu
}^{\alpha }$, other geometrical objects such as the components of the
curvature tensor $\widehat{R}_{\;\mu \beta \nu }^{\alpha }=$ $R_{\;\mu \beta
\nu }^{\alpha }=\Gamma _{\beta \mu ,\nu }^{\alpha }-\Gamma _{\mu \nu ,\beta
}^{\alpha }+\Gamma _{\rho \nu }^{\alpha }\Gamma _{\beta \mu }^{\rho }-\Gamma
_{\rho \beta }^{\alpha }\Gamma _{\nu \mu }^{\rho }$ , the components of the
Ricci tensor $\widehat{R}_{\mu \nu }=$ $R_{\mu \nu }=R_{\;\mu \alpha \nu
}^{\alpha }$ , the scalar curvature $\widehat{R}=\widehat{g}^{\mu \nu }%
\widehat{R}_{\mu \nu }=$ $\widehat{g}^{\mu \nu }R_{\mu \nu }=e^{\phi }g^{\mu
\nu }R_{\mu \nu }=e^{\phi }R$ are evidently invariant. Moreover, in a Weyl
integrable manifold it would be\ more natural to require this kind of
invariance to hold also in the definition of length, so we would redefine
the\ arc length of a curve $x^{\mu }=x^{\mu }(\lambda )$ between $x^{\mu }(a)
$ and $x^{\mu }(b)$ as 
\begin{equation}
\Delta s=\int_{a}^{b}\left( \widehat{g}_{\mu \nu }\frac{dx^{\mu }}{d\lambda }%
\frac{dx^{\nu }}{d\lambda }\right) ^{\frac{1}{2}}d\lambda =\int_{a}^{b}e^{-%
\frac{\phi }{2}}\left( g_{\mu \nu }\frac{dx^{\mu }}{d\lambda }\frac{dx^{\nu }%
}{d\lambda }\right) ^{\frac{1}{2}}d\lambda .  \label{weyl length}
\end{equation}

A second point concerns the interplay between covariant and contravariant
vectors in a Weyl integrable manifold. Let us examine how the isomorphism\
that exists between vectors and 1-forms is modified when the manifold is
endowed with an additional geometric field $\phi $. This question seems to
be relevant because, as we know, it is this duality that underlies the usual
operations of raising and lowering indices of vectors and tensors. In a Weyl
integrable manifold these operations make sense only if they fulfil the
requirement of Weyl invariance. Thus, let us now briefly recall how we show,
in the Riemannian context, that the tangent space $T_{p}(M)$ and the
cotangent space $T_{p}^{\ast }(M)$ at a point $p\in M$ are isomorphic \cite%
{Kerr}. The key point is to define the mapping $\widetilde{V}%
:T_{p}(M)\rightarrow 
\mathbb{R}
$ with $\widetilde{V}(U)=g(U,V)$ for any $U\in T_{p}(M)$. It is not
difficult to see that $\widetilde{V}$ is a 1-form and that to any 1-form $%
\sigma \in T_{p}^{\ast }(M)$ there corresponds a unique vector $V\in T_{p}(M)
$ such that $\sigma (U)=g(U,V)$. Now, assuming that $\{e_{\mu }\}$ and $%
\{e^{\mu }\}$ constitute dual bases for $T_{p}(M)$ and $T_{p}^{\ast }(M)$,
respectively,\ and putting $V=V^{\mu }e_{\mu }$ , $\sigma =\sigma _{\nu
}e^{\nu }$, we then have $\sigma _{\mu }=\sigma (e_{\mu })=g(e_{\mu
},V)=V^{\nu }g(e_{\nu },e_{\mu }).$ In view of the fact that $\sigma $ and $V
$ are isomorphic it is natural "to lower" the index $V^{\mu }$ by defining $%
V_{\mu }\equiv \sigma _{\mu }=g_{\nu \mu }V^{\nu }$, with $g_{\nu \mu
}\equiv g(e_{\nu },e_{\mu })$. Of course this procedure is not invariant
under Weyl transformations since the effective metric $\widehat{g}=e^{-\phi
}g$ does not enter in any of the above operations. To remedy this situation
it suffices to redefine the above algebra by replacing the Riemannian scalar
product $g:T_{p}(M)\times T_{p}(M)\rightarrow $ $%
\mathbb{R}
$ by a new scalar product given by the bilinear form $\widehat{g}%
:T_{p}(M)\times T_{p}(M)\rightarrow $ $%
\mathbb{R}
$ with $\widehat{g}(U,V)=e^{-\phi }g(U,V).$ In this way the operations of
raising and lowering indices when carried out with $\widehat{g}$ are clearly
invariant under (\ref{conformal}) and (\ref{gauge}).

Let us finally conclude\ this section with a few historical comments on Weyl
gravitational theory. Weyl developed an entirely new geometrical framework
to formulate his theory, the main goal of which was to unify gravity and
electromagnetism. As is well known, although admirably ingenious, Weyl's
gravitational theory turned out to be unacceptable as a physical theory, as
was immediately realized by Einstein, who raised objections to the theory 
\cite{Pauli,Pais}. Einstein's argument was that in a non-integrable\ Weyl
geometry the existence of sharp spectral lines in the presence of an
electromagnetic field would not be possible since atomic clocks would depend
on their past history \cite{Pauli}. However, the variant of Weyl geometry
known as Weyl integrable geometry does not suffer from the drawback pointed
out by Einstein. Indeed, it is the integral $I(a,b)=\int_{a}^{b}\sigma (%
\frac{d}{d\lambda })d\lambda $ that is responsible for the difference
between the readings of two identical atomic clocks following different
paths. Because in Weyl integrable geometry $I(a,b)$ is not path-dependent
the theory has attracted the attention of many cosmologists in recent years
as a viable geometrical framework for gravity theories \cite%
{Novello,Novello1}.

\section{General Relativity and a New Kind of Invariance}

We have seen in the previous section that the Weyl compatibility condition (%
\ref{compatibility}) is preserved\ when we go from a frame $(M,g,\phi )$ to
another frame $(M,\overline{g},\overline{\phi })$ through the
transformations (\ref{conformal}) and (\ref{gauge}). This has the
consequence that the components $\Gamma _{\mu \nu }^{\alpha }$ of the affine
connection are invariant under Weyl transformations, which, in turn, implies
the invariance of the affine geodesics. Now, as is well known, geodesics%
\textit{\ }plays a fundamental role in general relativity (GR) as well as in
any metric theory of gravity. Indeed, an elegant aspect of the
geometrization of the gravitational field lies in the geodesics postulate,
i.e. the statement that light rays\ and particles moving under the influence
of gravity alone follow space-time geodesics. Therefore a great deal of
information about the motion of particles in a given space-time is promptly
available once one knows its geodesics. The fact that geodesics are
invariant under (\ref{conformal}) and (\ref{gauge}) and that Riemannian
geometry is a particular case of Weyl geometry (when $\sigma $ vanishes, or $%
\phi $ is constant) seems to suggest that it should be possible to express
general relativity in a more general geometrical setting, namely, one in
which the form of the field equations is also invariant under Weyl
transformations. In this section, we shall show that this is indeed
possible, and we shall proceed through the following steps. First, we shall
assume that the space-time manifold which represents the arena of physical
phenomena may be described by a Weyl integrable geometry, which means that
gravity will be described by two geometric entities: a metric and a scalar
field. The second step is to set up an action $S$ invariant under Weyl
transformations. We shall require that $S$ be chosen such that there exists
a unique frame in which it reduces to the Einstein-Hilbert action. The third
step consists of extending Einstein's geodesic postulate to arbitrary
frames, such that in the Riemann frame it should describe the motion of test
particles and light exactly in the same way as predicted by general
relativity. Finally, the fourth step is to define proper time in an
arbitrary frame. This definition should be invariant under Weyl
transformations and coincide with the definition of GR's proper time in the
Riemann frame. It turns out then that the simplest action that can be built
under these conditions is%
\begin{equation}
S=\int d^{4}x\sqrt{-g}e^{-\phi }\left\{ R+2\Lambda e^{-\phi }+\kappa
e^{-\phi }L_{m}\right\} ,
\end{equation}%
where $R$ denotes the scalar curvature defined in terms of the Weyl
connection, $\Lambda $ is the cosmological constant, $L_{m}$ stands for the
Lagrangian of the matter fields and $\kappa $ is the Einstein's constant 
\footnote{%
Throughout this paper we shall adopt the following convention in the
definition of the Riemann and Ricci tensors: $R_{\;\mu \beta \nu }^{\alpha
}=\Gamma _{\beta \mu ,\nu }^{\alpha }-\Gamma _{\mu \nu ,\beta }^{\alpha
}+\Gamma _{\rho \nu }^{\alpha }\Gamma _{\beta \mu }^{\rho }-\Gamma _{\rho
\beta }^{\alpha }\Gamma _{\nu \mu }^{\rho };$ $R_{\mu \nu }=R_{\;\mu \alpha
\nu }^{\alpha }.$ In this convention, we shall write the Einstein equations
as $R_{\mu \nu }-\frac{1}{2}Rg_{\mu \nu }-\Lambda g_{\mu \nu }=-\kappa
T_{\mu \nu },$ with $\kappa =\frac{8\pi G}{c4}$.}. In $n$-dimensions we
would have 
\begin{equation}
S_{n}=\int d^{n}x\sqrt{-g}e^{\left( 1-\frac{n}{2}\right) \phi }\left\{
R+2\Lambda e^{-\phi }+\kappa e^{-\phi }L_{m}\right\} .  \label{action1}
\end{equation}

In order to see that the above action is, in fact, invariant with respect
to\ Weyl transformations, we just need to recall that under (\ref{conformal}%
) and (\ref{gauge}) we have $\overline{g}^{\mu \nu }=e^{-f}g^{\mu \nu }$, $%
\sqrt{-\overline{g}}=e^{\frac{n}{2}f}\sqrt{-g}$, $\overline{R}_{\;\nu \alpha
\beta }^{\mu }=R_{\;\nu \alpha \beta }^{\mu },$ $\overline{R}_{\mu \nu
}=R_{\mu \nu },$ $\overline{R}=\overline{g}^{\alpha \beta }\overline{R}%
_{\alpha \beta }=e^{-f}g^{\alpha \beta }R_{\alpha \beta }=$ $e^{-f}R$. It
will be assumed that $L_{m}$ depends on $\phi ,$ $g_{\mu \nu }$ and the
matter fields, here \ generically denoted\ by $\xi $, its form being
obtained from the special theory of relativity through the prescription $%
\eta _{\mu \nu }\rightarrow e^{-\phi }g_{\mu \nu }$ and $\partial _{\mu
}\rightarrow \nabla _{\mu }$, where $\nabla _{\mu }$ denotes the covariant
derivative with respect to the Weyl affine connection. If we designate the
Lagrangian of the matter fields in special relativity by $L_{m}^{sr}=$ $%
L_{m}^{sr}(\eta ,\xi ,\partial \xi )$, then the form of $L_{m}$ will be
given by the rule $L_{m}(g,\phi ,\xi ,\nabla \xi )\equiv L_{m}^{sr}(e^{-\phi
}g,\xi ,\nabla \xi )$.\ As it can be easily seen, these rules also ensure
the invariance under Weyl transformations of part of the action that is
responsible for the coupling of matter with the gravitational field, and, at
the same time, reproduce the principle of minimal coupling adopted in
general relativity when we set $\phi =0$, that is, when we go to the Riemann
frame by a Weyl transformation.

We now turn our attention to the motion of test particles and light rays.
Here, our task is to extend GR's geodesic postulate in such a way to make it
invariant under Weyl transformations. The extension is straightforward and
may be stated as follows: if we represent parametrically a timelike curve as 
$x^{\mu }=x^{\mu }(\lambda )$, then this curve\ will correspond to the world
line of a particle free from all non-gravitational forces, passing through
the events $x^{\mu }(a)$ and $x^{\mu }(b)$, if and only if it extremizes the
functional 
\begin{equation}
\Delta \tau =\int_{a}^{b}e^{-\frac{\phi }{2}}\left( g_{\mu \nu }\frac{%
dx^{\mu }}{d\lambda }\frac{dx^{\nu }}{d\lambda }\right) ^{\frac{1}{2}%
}d\lambda ,  \label{propertime}
\end{equation}%
which is obtained from the special relativistic expression of proper time by
using the prescription $\eta _{\mu \nu }\rightarrow e^{-\phi }g_{\mu \nu }.$
Clearly, the right-hand side of this equation is invariant under Weyl
transformations and reduces to the known expression of the propertime in
general relativity in the Riemann frame. We take $\Delta \tau $, as given
above, as the extension to an arbitrary Weyl frame, of GR's clock
hypothesis, i.e. the assumption that $\Delta \tau $ measures the proper time
measured by a clock attached to the particle \cite{Mainwaring}.

It is not difficult to verify that the extremization condition of the
functional (\ref{propertime}) leads to the equations 
\begin{equation*}
\frac{d^{2}x^{\mu }}{d\lambda ^{2}}+\left( \left\{ _{\alpha \beta }^{\mu
}\right\} -\frac{1}{2}g^{\mu \nu }(g_{\alpha \nu }\phi _{,\beta }+g_{\beta
\nu }\phi _{,\alpha }-g_{\alpha \beta }\phi \,_{,\nu })\right) \frac{%
dx^{\alpha }}{d\lambda }\frac{dx^{\beta }}{d\lambda }=0,
\end{equation*}%
where $\left\{ _{\alpha \beta }^{\mu }\right\} $ denotes the Christoffel
symbols calculated with $g_{\mu \nu }$. Let us recall that in the derivation
of the above equations the parameter $\lambda $ has been choosen such that%
\begin{equation}
e^{-\phi }g_{\alpha \beta }\frac{dx^{\alpha }}{d\lambda }\frac{dx^{\beta }}{%
d\lambda }=K=const.  \label{constant}
\end{equation}%
along the curve, which, up to an affine transformation, permits the
identification of $\lambda $ with the proper time $\tau $. It turns out that
these equations are exactly those that yield the affine geodesics in a Weyl
integrable space-time, since they can be rewritten as 
\begin{equation}
\frac{d^{2}x^{\mu }}{d\tau ^{2}}+\Gamma _{\alpha \beta }^{\mu }\frac{%
dx^{\alpha }}{d\tau }\frac{dx^{\beta }}{d\tau }=0,  \label{Weylgeodesics}
\end{equation}%
where $\Gamma _{\alpha \beta }^{\mu }=\left\{ _{\alpha \beta }^{\mu
}\right\} -\frac{1}{2}g^{\mu \nu }(g_{\alpha \nu }\phi _{,\beta }+g_{\beta
\nu }\phi _{,\alpha }-g_{\alpha \beta }\phi \,_{,\nu })$, according to (\ref%
{Weyl connection1}), may be identified\ with the components of the Weyl
connection. Therefore, the extension of the geodesic postulate by requiring
that the functional (\ref{propertime}) be an extremum is equivalent to
postulating that the particle motion must follow affine geodesics defined by
the Weyl connection $\Gamma _{\alpha \beta }^{\mu }$. It will be noted that,
as a consequence of the Weyl compatibility condition (\ref{compatibility})
between the connection and the metric, (\ref{constant}) holds automatically
along any affine geodesic determined by (\ref{Weylgeodesics}). Because both
the connection components $\Gamma _{\alpha \beta }^{\mu }$ and the proper
time $\tau $ are invariant when we switch from one Weyl frame to the other,
the equations (\ref{Weylgeodesics}) are invariant\ under Weyl
transformations.\ 

As we know, the geodesic postulate not only makes a statement about the
motion of particles, but also regulates the propagation of light rays in
space-time. Because the path of light rays are null curves, one cannot use
the proper time as a parameter to describe them. In fact, light rays are
supposed to follow null affine geodesics, which cannot be defined in terms
of the functional (\ref{propertime}), but, instead, they must be
characterized by their behaviour with respect to parallel transport. We
shall extend this postulate by simply assuming that light rays follow Weyl
null affine geodesics.

It is well known that null geodesics are preserved under conformal
transformations, although one needs to reparametrize the curve in the new
gauge. In the case of Weyl transformations, null geodesics are also
invariant with no need of reparametrization, since, again, the connection
components $\Gamma _{\alpha \beta }^{\mu }$ do not change under (\ref%
{conformal}) and (\ref{gauge}), while the condition (\ref{constant}) is
obvioulsy not altered. As a consequence, the causal structure of space-time
remains unchanged in all Weyl frames. This seems to complete our program of
formulating general relativity in a geometrical setting that exhibits a new
kind of invariance, namely, that with respect to Weyl transformations 
\footnote{%
We found that,\ in \cite{Poulis}, a similar action, in the case of vacuum,
was obtained by using an argument based on the Palatini approach.}.

\section{General Relativity as a Scalar-Tensor Theory}

In the present formalism it is interesting to rewrite the action (\ref%
{action1}) in Riemannian terms. This is done by expressing the Weyl scalar
curvature $R$ in terms of the Riemannian scalar curvature $\widetilde{R}$
and the scalar field $\phi $, which gives 
\begin{equation}
R=\widetilde{R}-(n-1)\square \phi +\frac{(n-1)(n-2)}{4}g^{\mu \nu }\phi
_{,\mu }\phi _{,\nu }\text{ ,}  \label{r}
\end{equation}%
where $\square \phi $ denotes the Laplace-Beltrami operator.\ \ It is easily
shown that, by inserting $R$ as given by (\ref{r}) into Eq.~(\ref{action1})
and using Gauss' theorem to neglect divergence terms in the integral, one
obtains 
\begin{equation}
S_{n}=\int d^{n}x\sqrt{-g}e^{\left( 1-\frac{n}{2}\right) \phi }\left\{ 
\widetilde{R}+\omega g^{\mu \nu }\phi _{,\mu }\phi _{,\nu }+2\Lambda
e^{-\phi }+\kappa e^{-\phi }L_{m}\right\} ,  \label{action2}
\end{equation}%
where $\omega =\frac{(n-1)(2-n)}{4}$. For $n=4$ we have $\omega =-\frac{3}{2}
$ and the action becomes 
\begin{equation}
S=\int d^{4}x\sqrt{-g}e^{-\phi }\left\{ \widetilde{R}-\frac{3}{2}g^{\mu \nu
}\phi _{,\mu }\phi _{,\nu }+2\Lambda e^{-\phi }+\kappa e^{-\phi
}L_{m}\right\} .  \label{action3}
\end{equation}

In the next section, it will be convenient to change the \ scalar field
variable $\phi $ by defining $\Phi =e^{-\phi }$. In terms of the new field $%
\Phi $, the action (\ref{action3}) takes the form 
\begin{equation}
S=\int d^{4}x\sqrt{-g}\left\{ \Phi \widetilde{R}-\frac{3}{2\Phi }g^{\mu \nu
}\Phi _{,\mu }\Phi _{,\nu }+2\Lambda \Phi ^{2}+\kappa \Phi ^{2}L_{m}\right\}
.  \label{action4}
\end{equation}

If we take variations of $S$, as given by (\ref{action3}), with respect to $%
g_{\mu \nu }$ and $\phi $, these being considered\ as independent fields, we
shall obtain, respectively, 
\begin{equation}
\widetilde{G}_{\mu \nu }-\phi _{,\mu ;\nu }+g_{\mu \nu }\square \phi -\frac{1%
}{2}(\phi _{,\mu }\phi _{,\nu }+\frac{1}{2}g_{\mu \nu }\phi _{,\alpha }\phi
^{,\alpha })=e^{-\phi }\Lambda g_{\mu \nu }-\kappa T_{\mu \nu },  \label{g}
\end{equation}%
\begin{equation}
\widetilde{R}-3\square \phi +\frac{3}{2}\phi _{,\alpha }\phi ^{,\alpha
}=\kappa T-4e^{-\phi }\Lambda ,  \label{fi}
\end{equation}%
where $\widetilde{G}_{\mu \nu }$ and $\widetilde{R}$\ denotes the Einstein
tensor and the curvature scalar, both calculated with the Riemannian
connection, and $T=g^{\mu \nu }T_{\mu \nu }.$\ It should be noted that (\ref%
{fi}) is just the trace of (\ref{g}), and so, the above equations are not
independent. This is consistent with the fact that we have complete freedom
in the choice of the Weyl frame. It also means that $\phi $ may be viewed as
an arbitrary gauge function and not as a dynamical field.

It is straightforward to verify that in terms of the variable $\Phi
=e^{-\phi }$, the equations (\ref{g}) and (\ref{fi}) read 
\begin{equation}
\widetilde{G}_{\mu \nu }=-\kappa T_{\mu \nu }+\Lambda \Phi g_{\mu \nu }+%
\frac{3}{2\Phi ^{2}}(\Phi _{,\mu }\Phi _{,\nu }-\frac{1}{2}g_{\mu \nu }\Phi
_{,\alpha }\Phi ^{,\alpha })-\frac{1}{\Phi }(\Phi _{,\mu ;\nu }-g_{\mu \nu
}\square \Phi ),  \label{gr1}
\end{equation}%
\begin{equation}
\widetilde{R}+3\frac{\square \Phi }{\Phi }-\frac{3}{2\Phi ^{2}}\Phi
_{,\alpha }\Phi ^{,\alpha }=\kappa T-4\Phi \Lambda .  \label{gr2}
\end{equation}%
\qquad

Some considerations should be made on the form taken by the energy-momentum
tensor $T_{\mu \nu }$, which appears on the right-hand side of the equations
(\ref{g}) and (\ref{gr1}). Here, as well as in the previous development of
the formalism that leads to the formulation of general relativity in a Weyl
integrable manifold, we use the effective metric $\widehat{g}=e^{-\phi }g$
as a guide to ensure Weyl invariance. In this way, it is natural to define
the energy-momentum tensor $T_{\mu \nu }(\phi ,g,\xi ,\nabla \xi )$ of the
matter field $\xi $, in an arbitrary Weyl frame $(M,g,\phi )$, by the
formula 
\begin{equation}
\delta \int d^{4}x\sqrt{-g}e^{-2\phi }L_{m}(g,\phi ,\xi ,\nabla \xi )=\int
d^{4}x\sqrt{-g}e^{-2\phi }T_{\mu \nu }(\phi ,g,\xi ,\nabla \xi )\delta
(e^{\phi }g^{\mu \nu }),  \label{energy-momentum}
\end{equation}%
where the variation on the left-hand side must be carried out simultaneously
with respect to both $g_{\mu \nu }$ and $\phi .$ In order to see that the
above definition makes sense, first recall that $L_{m}(g,\phi ,\xi ,\nabla
\xi )$ is given by the prescription $\eta _{\mu \nu }\rightarrow e^{-\phi
}g_{\mu \nu }$ and $\partial _{\mu }\rightarrow \nabla _{\mu }$, where $%
\nabla _{\mu }$ denotes the covariant derivative with respect to the Weyl
affine connection. Let us recall here that $L_{m}(g,\phi ,\xi ,\nabla \xi
)\equiv L_{m}^{sr}(e^{-\phi }g,\xi ,\nabla \xi )$, where $L_{m}^{sr}$
denotes the Lagrangian of the field $\xi $ in flat Minkowski space-time.
Secondly, it should be clear that the left-hand side of the equation (\ref%
{energy-momentum}) can always\ be put in the same form of the right-hand
side of the same equation. This can easily be seen from the fact that $%
\delta L_{m}=\frac{\partial L_{m}}{\partial g^{\mu \nu }}\delta g^{\mu \nu }+%
\frac{\partial L_{m}}{\partial \phi }\delta \phi =\frac{\partial L_{m}}{%
\partial (e^{\phi }g^{\mu \nu })}\delta (e^{\phi }g^{\mu \nu })$ and that $%
\delta (\sqrt{-g}e^{-2\phi })=-\frac{1}{2}\sqrt{-g}e^{-3\phi }g_{\mu \nu
}\delta (e^{\phi }g^{\mu \nu }).$ Finally, it is clear that the definition
of $T_{\mu \nu }(\phi ,g,\xi ,\nabla \xi )$ given by (\ref{energy-momentum})
is invariant under the Weyl transformations (\ref{conformal}) and (\ref%
{gauge}).

We would like to conclude this section with a brief comment on the form that
the equation that expresses the energy-momentum conservation law takes in a
arbitrary Weyl frame. We start with the Einstein's equations written in the
Riemann frame $(M,\widehat{g},0)$: 
\begin{equation}
G_{\mu \nu }(\widehat{g},0)=-\kappa T_{\mu \nu }(\widehat{g},0).
\label{einstein equation}
\end{equation}%
Because $G_{\mu \nu }(\widehat{g},0)$ is divergenceless with respect to the
metric connection $\{_{\mu \nu }^{\alpha }\}_{\widehat{g}}=$ $\frac{1}{2}%
\widehat{g}^{\alpha \beta }[\widehat{g}_{\beta \mu ,\nu }+\widehat{g}_{\beta
\nu ,\mu }-\widehat{g}_{\mu \nu ,\beta }]$ it follows from (\ref{einstein
equation}) that 
\begin{equation}
\widehat{\nabla }_{\alpha }T_{\mu }^{\;\alpha }=\widehat{\nabla }_{\alpha }(%
\widehat{g}^{\alpha \nu }T_{\mu \nu }^{\;})=0,  \label{conservation}
\end{equation}%
where the symbol $\widehat{\nabla }_{\alpha }$ denotes the covariant
derivative defined by $\{_{\mu \nu }^{\alpha }\}_{\widehat{g}}.$ If we now
go to an arbitrary Weyl frame $(M,g=e^{\phi }\widehat{g},\phi )$, then a
straightforward calculation shows that (\ref{conservation}) takes the form 
\begin{equation}
\nabla _{\alpha }T_{\mu }^{\;\alpha }=T_{\mu }^{\;\alpha }\phi _{,\alpha }-%
\frac{1}{2}T\phi _{,\mu }\text{ ,}  \label{conservation 2}
\end{equation}%
where $T=g^{\alpha \beta }T_{\alpha \beta }$ and $\nabla _{\alpha }$ stands
for the covariant derivative defined by the metric connection calculated
with $g$.

At first sight, due to the presence of non-vanishing terms on the right-hand
side of (\ref{conservation 2})\ one may be led to think that in the Weyl
frame we have an apparent violation of the energy-momentum conservation law.
Nonetheless, we must remember that if one is not working in the Riemann
frame the Weyl scalar field $\phi $ is an essential part of the geometry and
necessarily should appear in any equation describing the behaviour of matter
in space-time. This explain the presence of $\phi $ coupled with $T_{\mu \nu
}$ in (\ref{conservation 2}). Note that if $\phi =const$ we recover the
familiar \ general-relativistic energy-momentum conservation equation.
Finally, it is not difficult to verify that the above equation is invariant
under the Weyl transformations (\ref{conformal}) and (\ref{gauge}).

\section{Similarities with Brans-Dicke theory}

We shall now take a look at some similarities between the Brans-Dicke theory
of gravity and general relativity, when the latter is expressed in the
formalism we have developed in the previous section. For this purpose, let
us recall that the field equations of Brans-Dicke theory of gravity may be
written in the form \cite{BD} 
\begin{equation}
\widetilde{G}_{\mu \nu }=-\frac{\kappa ^{\ast }}{\Phi }T_{\mu \nu }-\frac{%
\omega }{\Phi ^{2}}(\Phi _{,\mu }\Phi _{,\nu }-\frac{1}{2}g_{\mu \nu }\Phi
_{,\alpha }\Phi ^{,\alpha })-\frac{1}{\Phi }(\Phi _{,\mu ;\nu }-g_{\mu \nu
}\square \Phi ),  \label{BD1}
\end{equation}%
\begin{equation}
\widetilde{R}-2\omega \frac{\square \Phi }{\Phi }+\frac{\omega }{\Phi ^{2}}%
\Phi _{,\alpha }\Phi ^{,\alpha }=0,  \label{BD2}
\end{equation}%
where $\kappa ^{\ast }=\frac{8\pi }{c^{4}}$, and we are keeping the notation
of the previous section, in which $\widetilde{G}_{\mu \nu }$ and $\widetilde{%
R}$\ denotes the Einstein tensor and the curvature scalar calculated with
respect to the metric $g_{\mu \nu }.$ By combining (\ref{BD1}) and (\ref{BD2}%
) we can easily derive the equation 
\begin{equation}
\square \Phi =\frac{\kappa ^{\ast }T}{2\omega +3},  \label{BD3}
\end{equation}%
which is the most common form of the scalar field equation usually found in
the literature. The equation (\ref{BD3}), however, is not defined for $%
\omega =-\frac{3}{2}$, so for this value of  $\omega $ one has to use (\ref%
{BD2}) instead, which then, becomes%
\begin{equation}
\widetilde{R}+3\frac{\square \Phi }{\Phi }-\frac{3}{2\Phi ^{2}}\Phi
_{,\alpha }\Phi ^{,\alpha }=0.  \label{BD4}
\end{equation}%
On the other hand, the equation (\ref{BD1}) for $\omega =-\frac{3}{2}$ reads 
\begin{equation}
\widetilde{G}_{\mu \nu }=-\frac{\kappa ^{\ast }}{\Phi }T_{\mu \nu }+\frac{3}{%
2\Phi ^{2}}(\Phi _{,\mu }\Phi _{,\nu }-\frac{1}{2}g_{\mu \nu }\Phi _{,\alpha
}\Phi ^{,\alpha })-\frac{1}{\Phi }(\Phi _{,\mu ;\nu }-g_{\mu \nu }\square
\Phi ).  \label{BD5}
\end{equation}%
Now, if we take the trace of the (\ref{BD5}) with respect to $g_{\mu \nu }$\
we get 
\begin{equation}
\widetilde{R}+3\frac{\square \Phi }{\Phi }-\frac{3}{2\Phi ^{2}}\Phi
_{,\alpha }\Phi ^{,\alpha }=\frac{\kappa ^{\ast }}{\Phi }T.  \label{BD6}
\end{equation}%
Of course (\ref{BD4}) and (\ref{BD6}) are not compatible, unless $T=0$,
which, then, implies that when $\omega =-\frac{3}{2}$ the Brans-Dicke field
equations (\ref{BD1}) and (\ref{BD2}) cease to be independent, and the
system of differential equations for \ $g_{\mu \nu }$ and $\Phi $ becomes
undertermined. As a consequence, one may freely choose an arbitrary $\Phi $
and work out a solution for $g_{\mu \nu }$ from (\ref{BD5}). In particular,
one can set $\Phi =\Phi _{0}=const$, in which case (\ref{BD5}) becomes
formally identical to the Einstein equations constant with the gravitational
constant $G$ replaced by $\frac{1}{\Phi _{0}}$. At this point, it is
interesting to note that one gets the same result by means of the conformal
transformation $\overline{g}_{\mu \nu }=e^{-\Phi }g_{\mu \nu }$, since the
conformally transformed Einstein tensor $\overline{G}_{\mu \nu }$\ is given
by $\overline{G}_{\mu \nu }=$ $\widetilde{G}_{\mu \nu }-\frac{3}{2\Phi ^{2}}%
(\Phi _{,\mu }\Phi _{,\nu }-\frac{1}{2}g_{\mu \nu }\Phi _{,\alpha }\Phi
^{,\alpha })+\frac{1}{\Phi }(\Phi _{,\mu ;\nu }-g_{\mu \nu }\square \Phi ).$%
( It is curious that one could use this property to generate an infinite
class of Brans-Dicke theory for $w=-\frac{3}{2}$ from known solutions of the
Einstein equations.) This known mathematical fact is often interpreted in
the literature as representing a conformal equivalence between Brans-Dicke
gravity for $w=-3/2$ and general relativity \cite{Dabrowski, Dabrowski1}. It
will be noted, however, that, in spite of the amazing similarity of the
field equations, we are far from having a complete analogy between the two
theories. Indeed, when we turn to the motion of test particles, we
immediately realize that in the Brans-Dicke theory it is postulated that
these particles must follow Riemannian geodesics, whereas in the case of GR
formulated in a Weyl frame (or in the case of conformal relativity) these
must follow geodesics that are not Riemannian. In the next section, we shall
illustrate this point with a simple example taken from a known vacuum
solution of Brans-Dicke theory, namely, the O`Hanlon-Tupper vacuum solution 
\cite{O'Hanlon}.

\section{Brans-Dicke vacuum solutions for w=-3/2}

In the case of vacuum and vanishing cosmological constant, the equations (%
\ref{gr1}) and (\ref{gr2}) reduce to 
\begin{equation}
\widetilde{G}_{\mu \nu }=\frac{3}{2\Phi ^{2}}(\Phi _{,\mu }\Phi _{,\nu }-%
\frac{1}{2}g_{\mu \nu }\Phi _{,\alpha }\Phi ^{,\alpha })-\frac{1}{\Phi }%
(\Phi _{,\mu ;\nu }-g_{\mu \nu }\square \Phi ),  \label{BDGR}
\end{equation}%
\begin{equation*}
\widetilde{R}+3\frac{\square \Phi }{\Phi }-\frac{3}{2\Phi ^{2}}\Phi
_{,\alpha }\Phi ^{,\alpha }=0,
\end{equation*}%
respectively. As we have just mentioned, in this situation the equations of
general relativity in an arbitrary Weyl frame ((\ref{gr1}) and (\ref{gr2}))
are identical to those of Brans-Dicke theory ((\ref{BD1}) and (\ref{BD2}))
for $\omega =-\frac{3}{2}$, provided that we identify the Weyl scalar field\
with the Brans-Dicke scalar field.\ At this point, suppose we want to see
how a solution of the above equations, regarded as a vacuum solution of
Brans-Dicke theory for $\omega =-\frac{3}{2},$ would look like when
interpreted as a vacuum solution of general relativity in a certain Weyl
frame, where the Brans-Dicke scalar field $\Phi $ now plays the role of the
Weyl scalar field. We can take, for instance, the well known O`Hanlon-Tupper
model, which is a vacuum solution of Brans-Dicke field equations
corresponding to a homogeneous isotropic space-time with spatial flat
section ($k=0)$. In this model, the metric $g_{\mu \nu }$\ and the scalar
field $\Phi $ are given, respectively, by 
\begin{subequations}
\begin{equation}
ds^{2}=dt^{2}-A(t)^{2}(dr^{2}+r^{2}d\theta ^{2}+r^{2}\sin ^{2}\theta
d\varphi ^{2}),  \label{Robertson}
\end{equation}%
where $A(t)=A_{0}t^{p}$, $\Phi =\Phi _{0}t^{q}$, with $p=\frac{1}{3\omega +4}%
(\omega +1\pm \sqrt{(2\omega +3)/3}),$ and $q=\frac{1}{3\omega +4}(1\mp 
\sqrt{3(2\omega +3)}$ , $A_{0}$ and $\Phi _{0}$ being integration constants 
\cite{O'Hanlon}. For $w>-\frac{3}{2}$ this solution has a big bang
singularity as $t\rightarrow 0.$ When $\omega \rightarrow \infty $ it has
the limit $A(t)=A_{0}t^{\frac{1}{3}},$ $\Phi (t)=\Phi _{0}=const$,\ which is
identical to the Friedmann model for stiff matter equation of state \cite%
{Mimoso}, and so this solution does not go over the corresponding general
relativistic solution, i.e., Minkowski space-time \cite{Barros,Faraoni}. For 
$\omega =-\frac{3}{2}$ we have $A(t)=A_{0}t$ and $\Phi =\Phi _{0}t^{-2}$ 
\footnote{
O%
\'{}%
Hanlon-Tupper solution for $\omega =-\frac{3}{2}$\ is identical to the
cosmological model found by Singh and Shridhar for a radiation-filled
Roberton-Walker universe \cite{Singh}.}$.$ This represents a model in which
the so-called \textit{Dirac's hypothesis}\ does not hold, since the
Newtonian gravitational "constant", interpreted in Brans-Dicke theory as the
inverse of the scalar field ($G\propto 1/\Phi )$, decreases as the universe
expands \cite{Dirac1}. 

In order to interpret the O%
\'{}%
Hanlon-Tupper model in the light of a general relativistic picture, we start
by putting (\ref{Robertson}) in the conformally-flat form 
\end{subequations}
\begin{equation}
ds^{2}=e^{\Psi (\tau )}(d\tau ^{2}-dr^{2}+r^{2}d\theta ^{2}+r^{2}\sin
^{2}\theta d\varphi ^{2}),  \label{OHT2}
\end{equation}%
where we have made the coordinate transformation $t=e^{A_{0}\tau }$ and
defined $\Psi (\tau )=2(\tau +\ln A_{0})$. In terms of the new coordinate,
the Brans-Dicke scalar field is given by $\Phi =$ $\Phi
_{0}A_{0}^{2}e^{-\Psi (\tau )}.$ Regarding both $g_{\mu \nu }$ given by $\ $(%
\ref{OHT2})$\ $\ and $\Phi $ as describing the gravitational field in the
Weyl frame $(M,g,\Phi )$, we now want to know how they\ will appear in a
Riemann frame $(M,\widehat{g},\widehat{\Phi })$, that is, in a frame, where $%
\widehat{\Phi }$ is constant and, hence, the geometry is\ Riemannian.
Recalling that the general form of \ theWeyl transformations (\ref{conformal}%
) and (\ref{gauge}) in terms of the variables $\Phi =e^{-\phi }$ and $%
\overline{\Phi }=e^{-\overline{\phi }}$\ is given by%
\begin{equation}
\widehat{g}_{\mu \nu }=e^{f}g_{\mu \nu },  \label{Weyl 1}
\end{equation}%
\begin{equation}
\widehat{\Phi }=e^{-f}\Phi ,  \label{Weyl 2}
\end{equation}%
it is clear that the natural choice of $f$ that will turn $\Phi $ into a
constant is $f=$ $-\Psi (\tau )$. We thus are led to the Riemann frame $(M,%
\widehat{g}=\eta ,\widehat{\Phi }=\Phi _{0}A_{0}^{2})$, where $\eta $
denotes Minkowski metric. Therefore, we conclude that the O%
\'{}%
Hanlon-Tupper cosmological model, when regarded formally as a general
relativistic solution in the Weyl frame $(M,g,\Phi )$, is equivalent to
Minkowski space-time, whose geodesics consists of straight lines satisfying
the equations%
\begin{equation}
\frac{d^{2}x^{\mu }}{d\tau ^{2}}=0.  \label{Minkowski}
\end{equation}%
From the fact the affine geodesics are invariant under the Weyl
transformations (\ref{conformal}) and (\ref{gauge}), and since in the
Riemann frame $(M,\widehat{g}=\eta ,\widehat{\Phi }=\Phi _{0}A_{0}^{2})$ the
Weyl affine geodesics coincide with the metric geodesics, it is evident that
in the in the Weyl frame $(M,g,\Phi )$ the affine geodesics will also be
given by (\ref{Minkowski}).

As we have already pointed out, the formal equivalence exhibited above
between Brans-Dicke vacuum solutions for $w=-\frac{3}{2}$ and general
relativistic vacuum solutions expressed in an Weyl geometric setting is not
complete. The reason is that we have not taken into account an aspect that
is fundamental to any metric theory of gravity: how do we determine the
motion of test particles and light. Indeed, as we have mentioned earlier, in
the case of general relativity, the geodesic equations that governs the
motion of test particles and light in an arbitrary Weyl frame are
constructed with the affine connection coefficients, which explicitly
involves the Weyl scalar field, and are invariant under Weyl
transformations. Of course we have a different situation in the case of
Brans-Dicke theory, where, even in the presence of the scalar field, the
geodesics are defined by the Levi-Civita connection. Therefore, in the O%
\'{}%
Hanlon-Tupper model the geodesic motion of particles and light will not be
given by (\ref{Minkowski}). A short calculation shows that the Brans-Dicke
geodesic equations are 
\begin{equation*}
\frac{d^{2}x^{\mu }}{d\tau ^{2}}+\frac{d\Psi }{d\tau }\frac{dx^{\mu }}{d\tau 
}+\frac{e^{-\Psi }}{2}\Psi ^{,\mu }\text{ }=0.
\end{equation*}

To conclude this section, we would like to show how the formal equivalence
discussed above can be used to generate a whole class of\ vacuum solutions
of Brans-Dicke field equations for $\omega =-\frac{3}{2}$, which includes
the O%
\'{}%
Hanlon-Tupper model as a particular case. To do this, let us suppose that we
want to obtain a solution of the field equations (\ref{BDGR}) corresponding
to a homogeneous and isotropic spacetime. As we know, the most general form
of the metric of such spacetime may be written as 
\begin{equation}
ds^{2}=dt^{2}-\frac{A(t)^{2}}{1+\frac{kr^{2}}{4}}(dr^{2}+r^{2}d\theta
^{2}+r^{2}\sin ^{2}\theta d\varphi ^{2}),  \label{FRW}
\end{equation}%
where $k=0,\pm 1$ represents the curvature of the spatial sections. \ We now
regard (\ref{BDGR}) as\ the Einstein's field equations in the Weyl frame $%
(M,g,\Phi )$, so that we can go to the Riemann frame $(M,\overline{g},%
\overline{\Phi }=1)$ \ through the Weyl transformations (\ref{Weyl 1}) and (%
\ref{Weyl 2}) by choosing $f=\ln \Phi $. In the Riemann frame, the line
element corresponding to $\overline{g}$ will be 
\begin{equation}
ds^{2}=\Phi (t)dt^{2}-\frac{\Phi (t)A(t)^{2}}{1+\frac{kr^{2}}{4}}%
(dr^{2}+r^{2}d\theta ^{2}+r^{2}\sin ^{2}\theta d\varphi ^{2}).\text{ }
\label{FRW2}
\end{equation}%
Defining a new time coordinate $\overline{t}$ by $\Phi (t)^{1/2}dt=d%
\overline{t}$ and putting $\Phi (t(\overline{t})A^{2}t(\overline{t})=%
\overline{A}^{2}(\overline{t})$,  (\ref{FRW}) takes the form

\begin{equation}
ds^{2}=d\overline{t}^{2}-\frac{\overline{A}(\overline{t})^{2}}{1+\frac{kr^{2}%
}{4}}(dr^{2}+r^{2}d\theta ^{2}+r^{2}\sin ^{2}\theta d\varphi ^{2})\text{.}
\label{FRW3}
\end{equation}%
Now, in the Riemann frame (\ref{BDGR}) becomes simply 
\begin{equation*}
\overline{G}_{\mu \nu }=0\text{ },
\end{equation*}%
with $\overline{G}_{\mu \nu }$ calculated with the metric $\overline{g}$.\
It may be readily verified that this yields only one independent equation,
namely, 
\begin{equation}
\left( \frac{d\overline{A}}{d\overline{t}}\right) ^{2}=-kc^{2}\text{ .}
\label{Friedmann}
\end{equation}%
An obvious conclusion that can be drawn from the above equation is that
there are no solutions for $k=1$ (this has been pointed out in (\cite%
{Dabrowski1}). If we take $k=0$, then $\overline{A}(\overline{t})=B$, where $%
B$ is an arbitrary constant. Thus, \ from the definition of \ $\overline{A}(%
\overline{t})$, we have $\Phi (t)A(t)^{2}=B$. \ This means that we have an
infinite number of Brans-Dicke vacuum solutions for $\omega =-\frac{3}{2}$, O%
\'{}%
Hanlon-Tupper model merely corresponding to\ the particular choice $A(t)\sim
t$.

\section{The Newtonian limit in a general Weyl frame}

In order to gain some insight into the meaning of this new representation of
general relativity developed in the previous sections, let us now proceed to
examine the Newtonian limit of general relativity in an arbitrary Weyl frame 
$(M,g,\phi )$.

As we know, a metric theory of gravity is said to possess a Newtonian limit
in the non-relativistic weak-field regime if one can derive\ from it
Newton's second law from the geodesic equations as well as the Poisson
equation from the gravitational field equations. Let us see how general
relatity when expressed in a form that is invariant under Weyl
transformations fulfills these requirements. The method we shall employ here
to treat this problem is standard and can be found in most textbooks on
general relativity ( see, for instance, \cite{Adler} ).

Since in Newtonian mechanics the space geometry is Euclidean, a weak
gravitational field in a geometric theory of gravity should manifest itself
as a metric phenomenon\ through a slight perturbation of the Minkowskian
space-time metric. Thus we consider a time-independent metric tensor of the
form 
\begin{equation}
g_{\mu \nu }=\eta _{\mu \nu }+\epsilon h_{\mu \nu },
\label{quasi-Minkowskian}
\end{equation}%
\ where $n_{\mu \nu }$ is the Minkowski tensor, $\epsilon $ is a small
parameter and the term $\epsilon h_{\mu \nu }$ represents a very small
time-independent perturbation due to the presence of some matter
configuration. Because we are working in the non-relativistic regime we
shall suppose that the velocity $V$ of the particle along the geodesic is
much less then $c$, so that the paramenter $\beta =\frac{V}{c}$ will be
regarded as very small; hence in our calculations only first-order terms in $%
\epsilon $ and $\beta $ will be kept. The same kind of approximation will be
assumed with respect to the Weyl scalar field $\phi $, which will be
supposed to be static and small, i.e. of the same order as $\epsilon $, and
to emphasise this fact we shall write $\phi =\epsilon \varphi $, where $%
\varphi $ is a finite function. Adopting then usual Minkowskian coordinates
of special relativity we can write the line element defined by (\ref%
{quasi-Minkowskian}) as 
\begin{equation*}
ds^{2}=(dx^{0})^{2}-(dx^{1})^{2}-(dx^{2})^{2}-(dx^{3})^{2}-\epsilon h_{\mu
\nu }dx^{\mu }dx^{\nu },
\end{equation*}%
which leads, in our approximation, to 
\begin{equation}
\left( \frac{ds}{dt}\right) ^{2}\cong c^{2}(1+\epsilon h_{00})\text{ .}
\label{dsdt}
\end{equation}%
We shall now consider, in the same approximation, the geodesic equations 
\begin{equation}
\frac{d^{2}x^{\mu }}{d\tau ^{2}}+\Gamma _{\;\alpha \beta }^{\mu }\frac{%
dx^{\alpha }}{d\tau }\frac{dx^{\beta }}{d\tau }=0,  \label{geodesics}
\end{equation}%
recalling that the symbol $\Gamma _{\;\alpha \beta }^{\mu }$ designates the
components of the Weyl affine connection. From (\ref{Weylconnection}) it is
easy to verify that, to first order in $\epsilon ,$ we have 
\begin{equation}
\Gamma _{\;\mu \nu }^{\alpha }=\frac{\epsilon }{2}n^{\alpha \lambda
}[h_{\lambda \mu ,\nu }+h_{\lambda \nu ,\mu }-h_{\mu \nu ,\lambda }+n_{\mu
\nu }\varphi _{,\lambda }-n_{\lambda \mu }\varphi _{,\nu }-n_{\lambda \nu
}\varphi _{,\mu }]\text{ .}  \label{Weylconnection2}
\end{equation}%
It is not difficult to see that, unless $\mu =\nu =0$, the product $\Gamma
_{\;\alpha \beta }^{\mu }\frac{dx^{\alpha }}{ds}\frac{dx^{\beta }}{ds}$ is
of order $\epsilon \beta $ or higher. In this way, the geodesic equations (%
\ref{geodesics}) become, to first order in $\epsilon $ and $\beta $%
\begin{equation*}
\frac{d^{2}x^{\mu }}{ds^{2}}+\Gamma _{\;00}^{\mu }\left( \frac{dx^{0}}{ds}%
\right) ^{2}=0\text{ .}
\end{equation*}%
By taking into account (\ref{dsdt}) the above equations may be written as 
\begin{equation}
\frac{d^{2}x^{\mu }}{dt^{2}}+c^{2}\Gamma _{\;00}^{\mu }=0\text{ .}
\label{equation-of-motion}
\end{equation}%
Clearly for $\mu =0$ the equation (\ref{equation-of-motion}) reduces to an
identity. On the other hand, if $\mu $ is a spatial index, a simple
calculation yields $\Gamma _{\;00}^{i}=-\frac{\epsilon }{2}\eta ^{ij}\frac{%
\partial }{\partial x^{j}}(h_{00}-\varphi )$, hence the geodesic equation in
this approximation becomes, in three-dimensional vector notation, 
\begin{equation*}
\frac{d^{2}\overrightarrow{X}}{dt^{2}}=-\frac{\epsilon }{2}c^{2}%
\overrightarrow{\nabla }(h_{00}-\varphi ),
\end{equation*}%
which is simply Newton's equation of motion in a classical gravitational
field provided we identify the scalar gravitational potential with 
\begin{equation}
U=\frac{\epsilon c^{2}}{2}(h_{00}-\varphi )\text{.}
\label{Newtonian-potential}
\end{equation}%
It is worth noting the presence of the Weyl field $\varphi $ in the above
equation. In fact, it is the combination $h_{00}-\varphi $ that represents
the\ Newtonian potential.

Let us now turn our attention to the Newtonian limit of the field equations.
For this purpose it will be convenient to recast the equation (\ref{g}) with 
$\Lambda =0$ into the form 
\begin{equation}
R_{\mu \nu }=-\kappa T_{\mu \nu }+\frac{1}{2}g_{\mu \nu }\left( \kappa
T+\square \phi \text{ }-\phi ,_{\alpha }\phi ^{,\alpha }\right) +\phi _{;\mu
;\nu }+\frac{1}{2}\phi _{,\mu }\phi _{,\nu }.  \label{Weyl-equations2}
\end{equation}%
In the weak-field approximation, i.e. when $g_{\mu \nu }=\eta _{\mu \nu
}+\epsilon h_{\mu \nu },$ it is easy to show that to first order in $%
\epsilon $, we have $R_{00}=$ $-\frac{1}{2}\nabla ^{2}\epsilon h_{00}$,
where $\nabla ^{2}$ denotes the Laplacian operator in flat space-time. On
the other hand, because we are assuming a static regime $\phi _{,0}=0$, so
the equation (\ref{Weyl-equations2}) for $\mu =\nu =0$ now reads 
\begin{equation}
\nabla ^{2}\left[ \frac{\epsilon c^{2}}{2}(h_{00}-\varphi )\right] =\kappa
(T_{00}-T).  \label{Newton-Weyl}
\end{equation}%
Let us consider a configuration of matter distribution with low proper
density $\rho $ moving at non-relativistic speed. The energy-momentum tensor
in this case is obtainable from special relativistic matter tensor 
\begin{equation}
T_{\mu \nu }=(\rho c^{2}+p)V_{\mu }V_{\nu }-p\eta _{\mu \nu },
\label{energy}
\end{equation}%
where $\rho $, $p$ and $V^{\mu }$ denotes, respectively, the proper density,
pressure and velocity field. We now need the expression of $T_{\mu \nu }$ in
an arbitrary Weyl frame. Rewriting this expression as $T_{\mu \nu }=(\rho
c^{2}+p)\eta _{\mu \alpha }\eta _{\nu \gamma }V^{\alpha }V^{\gamma }-p\eta
_{\mu \nu }$ and following the prescription $\eta _{\mu \nu }\rightarrow
e^{-\phi }g_{\mu \nu }$, we obtain 
\begin{equation}
T_{\mu \nu }=(\rho c^{2}+p)e^{-2\phi }g_{\mu \alpha }g_{\nu \gamma
}V^{\alpha }V^{\gamma }-pe^{-\phi }g_{\mu \nu },  \label{energy weyl}
\end{equation}%
which is the desired expression of the energy-momentum tensor in an
arbitrary frame. It is worth noting that in going from (\ref{energy}) to (%
\ref{energy weyl}) the quantities $\rho ,$ $p$ and $V^{\alpha }=\frac{%
dx^{\alpha }}{d\tau }$ are kept unaltered as, by definition, they are
invariant under Weyl transformations. Putting $e^{-\phi }\simeq 1-\epsilon
\varphi $ and recalling that in a non-relativistic regime we\ can neglect $p$
with respect to $\rho $, leads to $T_{00}=$ $T\simeq \rho c^{2}$. In this
way, we obtain, to first order in $\epsilon $ 
\begin{equation*}
T_{\mu \nu }\simeq \rho c^{2}\eta _{\mu \alpha }\eta _{\nu \gamma }V^{\alpha
}V^{\gamma }.\text{ }
\end{equation*}%
Finally,\ after substituting $\kappa =\frac{8\pi G}{c^{4}}$\ into the Eq. (%
\ref{Newton-Weyl}) we obtain 
\begin{equation}
\nabla ^{2}\left[ \frac{\epsilon c^{2}}{2}(h_{00}-\varphi )\right] =4\pi
G\rho ,  \label{A}
\end{equation}%
\ which clearly corresponds to the Poisson equation for the gravitational
field \ $\nabla ^{2}U=4\pi G\rho $ with $U$ given by (\ref%
{Newtonian-potential}).

\section{Different pictures of the same physical phenomena}

As we have seen, when we go from one frame\ $(M,g,\phi )$\ to another frame $%
(M,\overline{g},\overline{\phi })$ through the Weyl transformations (\ref%
{conformal}) and (\ref{gauge}), the pattern of affine\ geodesic curves does
not change. However, \ distinct geometrical and physical pictures may arise
in different frames. This is particular evident in the case of a conformally
flat space-time, i.e. when we have in a Riemann frame $g=e^{\phi }\eta $.\
In such situations, one can completely gauge away the Riemannian curvature
by a frame transformation, thereby going to a frame in which one is left
with a geometrical scalar field in a Minkowski background \cite{Pucheu}.
This is well illustrated, for instance, when we consider\ the class of
Robertson-Walker\ (RW) space-times ($k=0,\pm 1$), which are known to be
conformally flat \cite{Ibson}.\ If we go to the Weyl frame $(M,\eta ,\phi )$
by means of a Weyl transformation we arrive at a new cosmological scenario
in which the Riemannian curvature ceases to determine the cosmic expansion
and other phenomena, these effects being now attributed to the sole action
of a scalar field living in flat space-time. There are many other examples
of how distinct physical interpretations of the same phenomena are possible
in different frames. By way of illustration, we shall consider, in this
section, how one would describe, in \ a general Weyl frame, an important
effect predicted by general relativity: the so-called gravitational spectral
shift.

Let us consider the gravitational field generated by a massive body, which
in an arbitrary Weyl frame $(M,g,\phi )$\ is described by both the metric
tensor $g_{\mu \nu }$ and the scalar field $\phi $. For the sake of
simplicity, let us restrict ourselves to the case of a static field, in
which neither $g_{\mu \nu }$ nor $\phi $ depends on time. Let us suppose
that a light wave is emitted on the body at a fixed point with spatial
coordinates $(r_{E},\theta _{E},\varphi _{E})$ and received by an observer
at fixed point $(r_{R},\theta _{R},\varphi _{R}).$ Denoting the coordinate
times of emission and reception by $t_{E}$ and $t_{R}$, respectively, the
light signal, which in the Weyl frame corresponds to a null affine geodesic,
connects the event $(t_{E},r_{E},\theta _{E},\varphi _{E})$ with the event $%
(t_{R},r_{R},\theta _{R},\varphi _{R}).$ Let $\lambda $ be an affine
parameter along this null geodesic with $\lambda =\lambda _{E}$ at the event
of emission and $\lambda =\lambda _{R}$ at the event of reception. If we
write the line element in the form $ds^{2}=g_{00}(r,\theta ,\varphi
)dt^{2}-g_{jk}(r,\theta ,\varphi )dx^{j}dx^{k}$, then, since the geodesic is
null, we must have 
\begin{equation}
g_{00}(r,\theta ,\varphi )\left( \frac{dt}{d\lambda }\right)
^{2}=g_{jk}(r,\theta ,\varphi )\frac{dx^{j}}{d\lambda }\frac{dx^{k}}{%
d\lambda },  \label{spectralshift}
\end{equation}%
so we can write 
\begin{equation*}
\frac{dt}{d\lambda }=\left[ \frac{g_{jk}(r,\theta ,\varphi )}{%
g_{00}(r,\theta ,\varphi )}\frac{dx^{j}}{d\lambda }\frac{dx^{k}}{d\lambda }%
\right] ^{\frac{1}{2}}.
\end{equation*}%
On integrating between $\lambda =\lambda _{E}$ and $\lambda =\lambda _{R}$
we have 
\begin{equation}
t_{R}-t_{E}=\int \left[ \frac{g_{jk}(r,\theta ,\varphi )}{g_{00}(r,\theta
,\varphi )}\frac{dx^{j}}{d\lambda }\frac{dx^{k}}{d\lambda }\right] ^{\frac{1%
}{2}}d\lambda \text{ .}  \label{spectral2}
\end{equation}%
Because the integral on the right-hand side of the above equation depends
only on the light path through space, and since the emitter and observer are
at fixed positions in space, then $t_{R}-t_{E}$ has the same value for all
signals sent. This implies that for any two signals emmited at coordinate
times $t_{E}^{(1)},t_{E}^{(2)}$ and received at $t_{R}^{(1)},t_{R}^{(2)}$,
we have $t_{R}^{(1)}-$ $t_{E}^{(1)}=t_{R}^{(2)}-t_{E}^{(2)}$, which means
that the coordinate time difference $\Delta t_{E}=t_{E}^{(2)}-$ $t_{E}^{(1)}$
at the event of emission is equal to the coordinate time difference\ $\Delta
t_{R}=t_{R}^{(2)}-$ $t_{R}^{(1)}$\ at the event of reception. On the other
hand, we know from Section 3 that the proper time recorded by clocks in a
general Weyl frame must be calculated by using the formula 
\begin{equation*}
\Delta \tau =\int_{a}^{b}e^{-\frac{\phi }{2}}\left( g_{\mu \nu }\frac{%
dx^{\mu }}{d\lambda }\frac{dx^{\nu }}{d\lambda }\right) ^{\frac{1}{2}%
}d\lambda .
\end{equation*}%
Therefore, the proper time recorded by the clocks of observers situated at
the body and at the point of reception will be given, by the 
\begin{equation*}
\Delta \tau _{E}=e^{-\frac{\phi _{E}}{2}}\sqrt{g_{00}(r_{E},\theta
_{E},\varphi _{E})}\Delta t_{E},
\end{equation*}%
and 
\begin{equation*}
\Delta \tau _{R}=e^{-\frac{\phi _{R}}{2}}\sqrt{g_{00}(r_{R},\theta
_{R},\varphi _{R})}\Delta t_{R}.
\end{equation*}%
where $\phi _{E}=\phi (r_{E},\theta _{E},\varphi _{E})$ and $\phi _{R}=\phi
(r_{R},\theta _{R},\varphi _{R})$. Since $\Delta t_{E}=\Delta t_{R}$, we
have 
\begin{equation*}
\frac{\Delta \tau _{R}}{\Delta \tau _{E}}=\frac{e^{-\frac{\phi _{R}}{2}}%
\sqrt{g_{00}(r_{R},\theta _{R},\varphi _{R})}}{e^{-\frac{\phi _{E}}{2}}\sqrt{%
g_{00}(r_{E},\theta _{E},\varphi _{E})}}.
\end{equation*}%
Suppose now that $n$ waves of frequency $\nu _{E}$\bigskip\ are emitted in
proper time $\Delta \tau _{E}$ from an atom situated on the body. Then $\nu
_{E}=\frac{n}{\Delta \tau _{E}}$ is the proper frequency measured by an
observer situated at the body. On the other hand, the observer situated at
the fixed point $(r_{R},\theta _{R},\varphi _{R})$ will see these $n$ waves
in a proper time $\Delta \tau _{R}$, hence will measure a frequency $\nu
_{R}=\frac{n}{\Delta \tau _{R}}$. Therefore, we have 
\begin{equation}
\frac{\nu _{R}}{\nu _{E}}=\frac{e^{-\frac{\phi _{E}}{2}}\sqrt{%
g_{00}(r_{E},\theta _{E},\varphi _{E})}}{e^{-\frac{\phi _{R}}{2}}\sqrt{%
g_{00}(r_{R},\theta _{R},\varphi _{R})}}.  \label{spectralshiftweyl}
\end{equation}%
We, thus, see that $\nu _{R}\neq \nu _{E}$, i.e. the observed frequency
differs from the frequency measured at the body, and this constitutes the
spectral shift effect in a general Weyl frame.

To conclude, two points related to the above equation are worth noting. The
first is that since in a Riemann frame $\phi =0$ the Eq. (\ref%
{spectralshiftweyl}) reduces the well-known general relativistic formula for
the gravitational spectral shift. The second point is that if we go to a
Weyl frame where $g_{00}$ is constant, then Eq. (\ref{spectralshiftweyl})
becomes simply 
\begin{equation*}
\frac{\nu _{R}}{\nu _{E}}=e^{\frac{1}{2}(\phi _{R}-\phi _{E})}.
\end{equation*}%
As we see, in this frame all information concerning the gravitational field
is contained in the Weyl scalar field.

\section{WIST theory viewed in the Riemann frame}

In Section 2, we have briefly commented on the close correspondence between
the mathematical structure of Weyl integral geometry and Riemmanian
geometry. More precisely, we have shown that to each Weyl frame $(M,g,\phi )$%
\ there corresponds a unique Riemann frame $(M,\widehat{g}=e^{-\phi }g,0),$
such that geometrical objects constructed from $g$ and $\phi $ in the frame $%
(M,g,\phi )$, such as the affine connection coefficients, curvature,
geodesics, etc, can be carried over to $(M,\widehat{g},0)$ without
ambiguity, and vice-versa. This fact makes us wonder how some gravity
theories formulated in a Weyl integral space-time would then appear when
viewed in the Riemann frame $(M,\widehat{g}=e^{-\phi }g,0)$. A good
representative of these theories, in which we would like to focus our
attention now, is a proposal known as the \textit{Weyl integrable space-time 
}(WIST) \cite{Novello}. Let us recall the basic tenets of this theory.

The WIST approach starts by postulating the action 
\begin{equation}
S=\int d^{4}x\sqrt{-g}\left\{ R+\omega \phi ^{,\mu }\phi _{,\mu }+e^{-2\phi
}L_{m}\right\} ,  \label{action wist}
\end{equation}%
where $R$ denotes the Weylian curvature, $\phi $ is the scalar Weyl field, $%
\omega $ is a dimensionless parameter and $L_{m}$ is the Lagrangian of the
matter fields. It is also postulated that the form of $L_{m}$ is obtained
from the corresponding Lagrangian in special relativity by substituting
simple derivatives by covariant derivatives with respect to the Weyl
connection. As regards to the above action, two comments are in order. The
first is that it is not invariant under the Weyl transformations (\ref%
{conformal}) and (\ref{gauge}). The second, as we shall show now, is that
when we go to the Riemann frame $(M,\widehat{g}=e^{-\phi }g,0)$ through the
Weyl transformations $\widehat{g}_{\mu \nu }=e^{-\phi }g_{\mu \nu },\widehat{%
\phi }=\phi -\phi =0$, then (\ref{action wist}) becomes 
\begin{equation}
S=\int d^{4}x\sqrt{-\widehat{g}}e^{\phi }\left\{ \widehat{R}+\omega \widehat{%
g}^{\mu \nu }\phi _{,\mu }\phi _{,\nu }+L_{m}\right\} ,
\label{action wist 2}
\end{equation}%
where by $\widehat{R}$ we are denoting the scalar curvature defined in terms
of $\widehat{g}_{\mu \nu }.$ Changing to the field variable $\Phi =e^{\phi }$%
, we finally get 
\begin{equation}
S=\int d^{4}x\sqrt{-\widehat{g}}\left\{ \Phi \widehat{R}+\frac{\omega }{\Phi 
}\widehat{g}^{\mu \nu }\Phi _{,\mu }\Phi _{,\nu }+L_{m}\right\} ,
\label{BD action}
\end{equation}%
which we immediately recognise as the action of Brans-Dicke theory of
gravity written in units such that $\frac{8\pi }{c^{4}}=1$ \cite{BD}$.$ We,
thus, see that in \ the Riemann frame the WIST action (\ref{action wist}) is
formally identical to the Brans-Dicke action (\ref{BD action}), where $\Phi $
is no longer interpreted as a geometrical field. This reminds us of a
similar situation in which Brans-Dicke theory is interpreted in two
different frames, the Jordan and Einstein frames, an issue widely discussed
in the literature \cite{Faraoni2}.

The mathematical analogy between WIST and Brans-Dicke theories works in both
directions. Thus, one may start the action (\ref{BD action}), which gives
Brans-Dicke theory in the usual Riemannian (Jordan) frame, and then go to
the Weyl frame (Einstein frame) in which the action takes the form of (\ref%
{action wist}), where the scalar field $\phi $ might be interpreted as a
geometric field.\ The usual view, let us say, the non-geometrical view, is
that we have the same Brans-Dicke theory in two different frames, the Jordan
and Einstein frame. The physical interpretation of the two pictures has been
widely discussed in the literature \cite{Faraoni2}.\ However, a
characteristic feature of Brans-Dicke theory is that Newton's gravitational
constant $G$  is replaced by the inverse of the scalar field, i.e. $G=$ $%
\Phi ^{-1}$, an idea that goes back to Dirac \cite{Dirac1}. Similarly to the
original Weyl theory, which represents an elegant way of geometrizing the
electromagnetic field \cite{Weyl}, the same can be said of the WIST theory
as regards to the scalar field: we have here a geometrization of a scalar
field. In view of this analogy, the passage from the Jordan frame to the
Einstein frame may be interpreted as a \ "geometrization" of $G$, the
empirical physical quantity that sets the strength of the gravitational
force, now promoted to the status of a field. One may perhaps feel inclined
to regard this geometrical attempt to explain the origin of $G$ as\ being in
accordance with the Machian view that local physical laws are determined by
the large-scale structure (geometry) of the universe \cite{Hawking}.

It is worth noting that a connection between Brans-Dicke theory and Weyl
integrable geometry appears in a different context. In fact, this connection
has been proved to exist for any scalar-tensor theory in which the scalar
field is non-minimally coupled to the metric \cite{Novello 2,Poulis} .
Without going into the details, the argument is the following. We start with
the action (\ref{action wist 2}) in the absence of matter and consider
variations in the sense of Palatini approach, i.e. treating the metric and
the affine connection separately as dynamical variables. It is then not
difficult to show that the variation with respect to the connection leads to
the equation (\ref{compatibility}), that is, the compatibility condition
that defines a Weyl integrable manifold.

\section{Final remarks}

As we have seen, it is possible to set up a different scenario of general
relativity theory in which the gravitational field is not associated with
the metric tensor only, but with the combination of both the metric $g_{\mu
\nu }$ and a geometrical scalar field $\phi $. In this scenario we have a
new kind of invariance and\ the same physical phenomena may appear in
different pictures and distinct representations. This can be well
illustrated when we consider, for instance, homogeneous and isotropic
cosmological models. All these have a conformally-flat geometry, and as a
consequence, there is a frame in which the geometry of these models becomes
that of flat Minkowski space-time. In the Riemann frame the space-time
manifold is endowed with a metric that leads to Riemannian curvature, while
in the Weyl frame space-time is flat. In this case, all information about
the gravitational field is encoded in the scalar field. Another example is
given by the gravitational spectral shift, in which the Weyl scalar field
plays an essential role.

The presence of a scalar field in an arbitrary Weyl frame also leads to
formal analogy with Brans-Dicke theory, a fact that has\ already been known
and mentioned in the literature \cite{Dabrowski}. Because of this
O`Hanlon-Tupper space-time in Brans-Dicke theory with $\omega =-\frac{3}{2}$
can be regarded as Minkowski space-time in a Weyl frame, although the
analogy is not perfect since in Brans-Dicke theory test particles follow
metric geodesics rather than affine Weyl geodesics.

An important conclusion to be drawn from what has been presented in this
paper is that general relativity can perfectly \textquotedblleft
survive\textquotedblright\ in a non-Riemannian environment. Moreover, as far
as physical observations are concerned, all Weyl frames, each one
determining a specific geometry, are completely equivalent. In a certain
sense, this\ would reminds us of the view conceived by H. Poincar\'{e} that
the geometry of space-time is perhaps a convention that can be freely chosen
by the theoretician \cite{Henri}. In particular, according to this view,
general relativity might be rewritten in terms an arbitrary conventional
geometry \cite{Tavakol}. 

Finally, we should also note that\ the same formalism we have used to recast
general relativity in a form that is manifestly invariant under Weyl
transformations may be extended in a straightforward way to the so-called $%
f(R)$ theories \cite{Sotiriou}, where the issue of physical interpretation
between the Einstein and Jordan frames may be of interest \cite{Brown}. The
basic idea here is to start with the action $S=\int d^{4}x\sqrt{-g}%
\{f(R)+\kappa L_{m}(g,\xi )\}$, where $\xi $ stands generically for the
matter fields. We then follow the same procedure presented in Section 3 and
postulate that this action may be regarded as defined in a Weyl integral
space-time in a particular frame where the Weyl scalar field vanishes, that
is, in the Riemann frame. The next step is almost obvious: using the fact
that the combination $e^{\phi }R$ is invariant under (\ref{conformal})\ and (%
\ref{gauge}) the sought-after action in an arbitrary Weyl frame will be
given by $S=\int d^{4}x\sqrt{-g}e^{-2\phi }\{f(e^{\phi }R)+\kappa
L_{m}(g,\xi )\}$, where for the definition of $L_{m}(g,\xi )$ in an
arbitrary frame the prescriptions outlined in Section 3 still apply. We
leave the details of this extension for a separate publication.

\section*{Acknowledgments}

C. Romero and M. L. Pucheu would like to thank CNPq/CLAF for financial
support. We are grateful to Dr. I. Lobo for helpful discussions and
suggestions.

\end{document}